\documentclass[amsmath,twocolumn,amssymb,amsbsy, titlepage]{revtex4-1}

\usepackage{graphicx}
\usepackage{nicefrac}
\usepackage{amsfonts}

\usepackage{amssymb}
\usepackage{amsmath} 

\usepackage{subfigure}
\usepackage{multirow} 
\usepackage{tabularx} 
\usepackage{array}

\usepackage{color}

\usepackage{units}

\usepackage{tensor} 
\usepackage{braket}

\usepackage{bm}
\usepackage{hyperref}

\begin{document}
\title{Boundary driven Heisenberg-chain in the long-range interacting regime: Robustness against far from equilibrium effects}
\author{Leon Droenner}
\email{droenner@itp.tu-berlin.de}
\author{Alexander Carmele} 
\affiliation{Institut f\"ur Theoretische Physik, Nichtlineare Optik und Quantenelektronik, Technische Universit\"at Berlin, 10623 Berlin, Germany}

\begin{abstract}
We investigate the Heisenberg XXZ-chain with long-range interactions in the Z-dimension.  By applying two magnetic boundary reservoirs we drive the system out of equilibrium and induce a non-zero steady state current. The long-range coupled chain shows nearly ballistic transport and linear response for all potential differences of the external reservoirs. In contrast, the common isotropic nearest-neighbor coupling  shows negative differential conductivity and a transition from diffusive to subdiffusive transport for a far from equilibrium driving. Adding disorder, the change in the transport for nearest neighbor coupling is therefore highly dependent on the driving. We find for the disordered long-range coupled XXZ-chain, any change in the transport behavior is independent of the potential difference and the coupling strengths of the external reservoirs.  

\end{abstract}

\maketitle
The study of generic strongly-correlated quantum spin-models is of fundamental importance to understand underlying quantum phase transitions. Recently it became experimentally accessible to tune the interactions between spins such that long-range interactions can be investigated \cite{NATUREVOL484, Islam583, NATUREVOL511, NATUREVOL511202, Neyenhuise1700672, PhysRevLett.119.080501}. Long-range interactions occur in real physical systems e.g. Coulomb interaction. A generalization of the well studied nearest-neighbor scenario to long-range interactions in spin-models provides a deeper understanding and has drawn a lot of interest in the recent years \cite{PhysRevLett.104.137204, PhysRevA.92.041603, PhysRevB.95.024431}. One example for a phase transition is the many-body localization (MBL) transition \cite{Basko_MBL,huse:MBL_strong_interaction, Thermal-eigenstates, MBL_integegrability, nandkishore/huse:MBL, MBL:renormalization_group, Experiment_MBL:Schreiber, Experiment_MBL:Bordia, 2015_MBL_quantum_simulator, PhysRevB.92.195107}, the  generalization of Anderson localization \cite{Anderson} for interacting systems, e.g. the disordered Heisenberg spin-chain. Recent publications provide a MBL transition in models with long-range interactions \cite{2015_MBL_quantum_simulator, PhysRevB.92.134204, PhysRevB.91.094202, PhysRevLett.113.243002}. While it has been demonstrated that MBL exists in an isolated quantum system as the Heisenberg XXZ-chain\cite{Znidaric:entanglement_DMRG, Mobility_edge_Heisenberg} it is still challenging in the presence of an external bath. Several approaches considered the existence of a thermalizing reservoir regarding the spectral properties \cite{spectral_features_MBL_coupling_to_bath1, spectral_features_MBL_coupling_to_bath2}, while others focused on the effect of dephasing referring to the measurement of optical lattice systems\cite{dephasing_MBL_purity, Coupled_to_a_bath_lindblad_MBL, dephasing_MBL_znidaric}.\\
One method to characterize MBL is the strictly zero conductivity in the localized phase, associated with a transition from metallic to insulating behavior \cite{Basko_MBL}. Thus, vanishing current at a certain disorder strength indicates the MBL transition. A well studied approach to an open Heisenberg model is connecting the system to two magnetic reservoirs at the boundaries \cite{0295-5075-61-1-034, PhysRevE.76.031115, PhysRevLett.106.220601, PhysRevLett.106.217206, PhysRevLett.110.047201, PhysRevE.88.062118, PhysRevB.91.174422, 1742-5468-2017-4-043302}. By keeping the reservoirs at different potentials, a spin current is induced. For nearest-neighbor coupling it has been demonstrated that such boundary driven spin-models provide non-equilibrium features such as negative differential conductivity (NDC) \cite{negative_differential_conductivity, driving_clean_chain} or anomalous transport \cite{ZNIDARIC_isotropic_2011, Prozen_transport}. Dephasing  enhances the transport \cite{Znidaric:dephasing, dephasing_spinless_fermions, 1742-5468-2013-07-P07007}. If disorder is applied, it remains an open question whether MBL survives in these boundary driven systems. Still, the ergodic side with weak disorder contains a transition from diffusive to subdiffusive transport \cite{diffusive/subdiffusive_Lindblad_chain}. For the isolated Heisenberg-chain different results were obtained with either a transition from diffusive transport to a subdiffusive regime close to the MBL transition \cite{PhysRevLett.114.160401, PhysRevB.94.180401} or a subdiffusive regime up until zero disorder \cite{PhysRevLett.114.100601, PhysRevB.93.060201, PhysRevB.93.224205}. We capture the idea of ref. \cite{diffusive/subdiffusive_Lindblad_chain} but study the transport properties of the system being far from equilibrium, finding a subdiffusive regime all the way to zero disorder.  Due to the far from equilibrium situation, the system builds up ferromagnetic domains at the boundaries (spin-blockade)\cite{driving_clean_chain} leading to NDC \cite{negative_differential_conductivity} which suppresses the transport already for zero disorder.  We find that for a long-range interaction for the Ising-part, the system is robust against far from equilibrium effects and NDC is absent. The transport remains nearly ballistic for all investigated potential differences and reservoir coupling strengths. In case of disorder, long-range coupling shows a transition to subdiffusive transport at a disorder strength independent of the external reservoirs. This indicates a great difference between long-range coupling and nearest neighbor coupling for the far from equilibrium situation.   Our findings propose that long-range coupling makes an ideal candidate to study many-body localization for the situation of two boundary reservoirs as the transport is not dependent on the bath induced localization at the boundaries.
\section{Model}\label{model}
\begin{figure}[h]
\center
\includegraphics[width=0.48\textwidth]{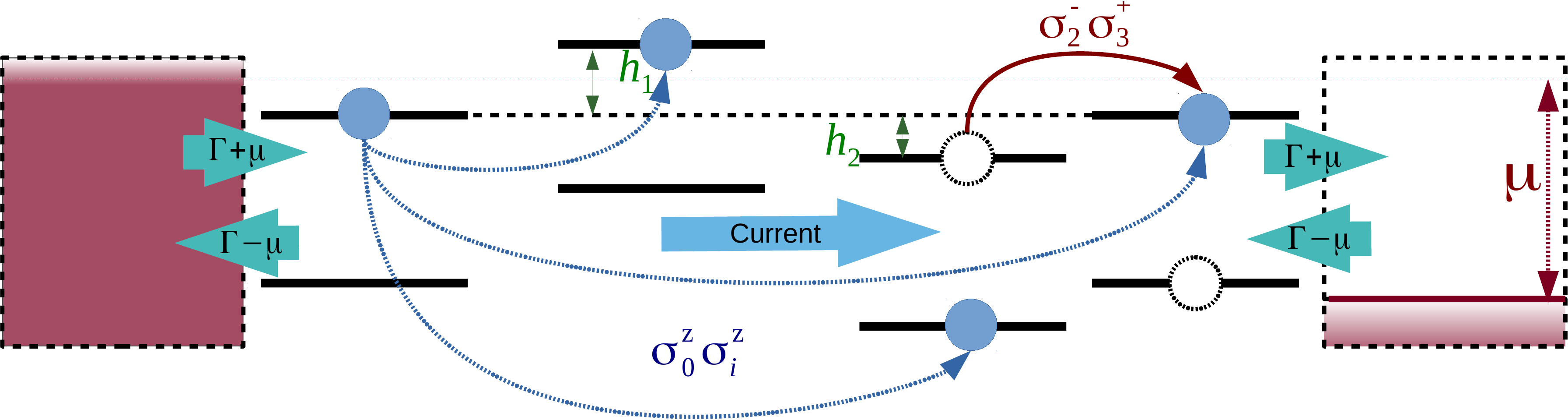}
\caption{Model of the Heisenberg-chain coupled to two Lindblad reservoirs at the boundaries: Neighboring spins can flip which is demonstrated as excitation hopping via $(\sigma^-_i\sigma^+_{i+1})^{(\dagger)}$. The interactive  Ising part $\sigma^z_i\sigma^z_{i+l}$ will be either a nearest-neighbor ($l=1$) or a long-range coupling. Two Lindblad reservoirs are applied at the edges of the chain with excitation in- and outscattering rates $\Gamma$. The chain is driven via a potential difference of the reservoirs $\mu$ which induces a spin current.}
\label{open_Heisenbergchain}
\end{figure}
We consider a Heisenberg quantum spin-chain consistent of $N$ spins. We focus on the isotropic case where all spin coupling constants $J$ are equal. We describe the Hamiltonian via Pauli spin-matrices $\left\{\sigma^x,\sigma^y,\sigma^z\right\}$, obtained from the Spin-operator $\boldsymbol S=\frac{\hbar}{2} \boldsymbol \sigma$. The Hamiltonian reads  ($\hbar=1$)
\begin{align}
\mathcal{H}=&\sum_{i=0}^{N-2}\frac{J}{4}\left( \sigma^x_i \sigma^x_{i+1}+  \sigma^y_i \sigma^y_{i+1}\right)\notag\\
&+\frac{1}{4}\frac{J}{A}\sum_{i=0}^{N-2}\sum^{N-1}_{l>i}\frac{1}{|l-i|^\alpha}\sigma^z_i\sigma^z_{l}+\sum^{N-1}_{i=0}\frac{h_i}{2}\sigma_i^z\label{Heisenbergchain}\\
A=&\frac{1}{(N-1)}\sum_{i=0}^{N-2} \sum_{l>i}^{N-1} \frac{1}{|l-i|^\alpha}\label{norm}~,
\end{align}
whereby the first two terms describe spin-flips between neighboring sites. The second term is an Ising-like interaction between the spins.\\
The Heisenberg spin-chain is equal to a model of spinless fermions\cite{driving_clean_chain}, which is why we illustrate the spins as two-level systems  with one fermion per site. Thus, the upper level is equal to spin up and and vice versa. Our model system is schematically shown in Fig. \ref{open_Heisenbergchain}.\\
We apply a long-range coupling within the interaction term $\sigma^z_i\sigma^z_{i+1}$ where $\alpha$ differs between the coupling scenarios. If $\alpha \rightarrow \infty$ it reproduces the nearest neighbor coupling, leading to the standard disordered isotropic Heisenberg spin-chain. We assume $\alpha=1000$ as the nearest-neighbor case. The case $\alpha \rightarrow 0$ describes equal coupling with all other sites. All other values of $\alpha$ represent a decaying coupling with the distance. We consider the case $\alpha=0.5$ as long-range coupling, where the interaction decays with the square root of the distance. In order to compare the two different coupling-scenarios, we normalize the  long-range coupling with weight on $\alpha$ as shown in Eq. \eqref{norm}. Thus, the limiting cases of $\alpha \rightarrow \infty$ yields $A=1$ and $\alpha=0$ defines $A=N/2$. \\
Within Sec. \ref{effect of disorder} we investigate the effect of disorder which describes the last term in Eq. \ref{Heisenbergchain}. We apply disorder via $h_i\in [-h,h] $ randomly at each site of the chain, shifting the on-site energies.\\
In order to induce and control spin transport, we apply two different reservoirs which act on the boundary spins $\{0,N-1\}$. We model the system-reservoir interaction with the Lindblad master equation and evaluate the full density-matrix dynamics via 
\begin{align}
\partial_t \rho(t)=-i\left[\mathcal{H},\rho(t)\right]+\sum_{\substack{ j\in \{L,R\} \\ k \in \{+,-\}  }} \mathcal{D}\left[L^j_k\right]\rho(t)~.\label{density-matrix-eq_with_driving}
\end{align}
The Lindblad-terms describe a generic system-reservoir interaction within Born-Markov and secular approximation. The typical Lindblad-form with superoperator $\mathcal{D}$ reads 
\begin{align}
\mathcal{D}[\hat x]\rho=2\hat x \rho \hat x^\dagger -\left\{\hat x^\dagger \hat x, \rho\right\}~,
\end{align}
with anticommutator brackets $\{a,a^\dagger\}=aa^\dagger+a^\dagger a$. At the left side of the chain we use for the operator  $\hat x$ the Lindbladians\cite{diffusive/subdiffusive_Lindblad_chain}
\begin{align}
&L^L_+=\frac{1}{2}\sqrt{\Gamma(1+\mu)}\sigma_0^+,~ L^L_-=\frac{1}{2}\sqrt{\Gamma(1-\mu)}\sigma_0^-
\label{Lindbladian}
\end{align}
and for the right side the complex conjugate
\begin{align}
&L^R_+=\frac{1}{2}\sqrt{\Gamma(1-\mu)}\sigma_{N-1}^+,~ L^R_-=\frac{1}{2}\sqrt{\Gamma(1+\mu)}\sigma_{N-1}^-~.
\label{Lindbladian_conjugate}
\end{align}
We introduce the driving parameter $\mu\geq 0$ phenomenologically as a potential difference between the two reservoirs as it can be seen in Fig. \ref{open_Heisenbergchain}. Due to the different in- and outscattering for the left and the right boundary spin, $\mu > 0$ induces a spin-current through the chain. For $\mu\ll 1$, the system acts within a linear response regime. Maximal driving is obtained for $\mu=1$ which describes just inscattering at the left and outscattering at the right side. Or in other words for the latter case, the left reservoir contains only spin-up magnetization while the right reservoir only includes magnetization of spin-down.
We initialize all sites with spin-down $\rho(0)=|\downarrow\downarrow\downarrow...\rangle\langle \downarrow\downarrow\downarrow...|$. Due to the boundary reservoirs at different potentials, the system converges to a non-equilibrium steady state.\\
We focus on the spin-current which is derived via the continuity equation
\begin{align}
-\dot{\sigma_k^z}=\frac{j_k-j_{k-1}}{k-(k-1)}=j_{k-1}-j_k~.
\end{align}
Calculating the time derivative with Heisenberg equation of motion $\dot{\sigma_k^z}=i\left[\sigma_k^z,H\right]$, the spin-current reads
\begin{align}
j_k=&\frac{J}{4}\left(\sigma_k^x\sigma^y_{k+1}-\sigma_k^y\sigma^x_{k+1}\right)~.\label{current}
\end{align}

Reaching the non-equilibrium steady state, the relative spin-current is independent of the site index and therefore $\left\langle j\right\rangle=\text{lim}_{t\rightarrow\infty}\text{Tr}\left(\rho(t)j_k\right)$.\\

\section{Characterizing Spin-transport}\label{characterizing spin-transport}
We investigate the transport of both models, either nearest-neighbor coupling ($\alpha=1000$) or long-range coupling ($\alpha=0.5$). Hereby we show that a long-range coupled chain still acts within a linear response regime for all kind of external driving in contrast to the nearest-neighbor scenario. For the following plots we set $J=\Gamma$.
\begin{figure}[h]

\centering
\includegraphics[width=0.48\textwidth]{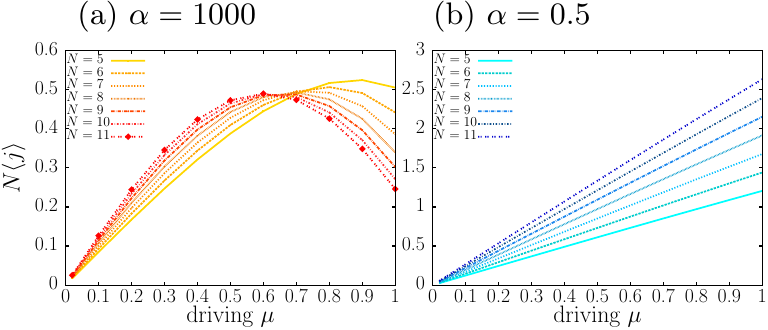}
\caption{Comparison between the nearest-neighbor scenario (a) $\alpha=1000$ and long-range coupling (b) $\alpha=0.5$ for different chain lengths. We show the absolute current $N\langle j\rangle$ versus the driving strength $\mu$.  For weak driving, both scenarios are within a linear response regime where the current increases linearly with the driving strength. For nearest-neighbor coupling (a), the current reaches a maximum at a certain driving strength which is dependent on the system size. After the maximum, the current decreases and shows NDC due to the building of a wide spin-blockade. At $\mu^{diff} \approx 0.6$ there is a crossing of the current for different system sizes signifying diffusive transport. In contrast, for long-range coupling (b) the system is still within a linear-response regime for maximal driving. }

\label{figure:driving}
\end{figure}
In Fig. \ref{figure:driving} we show the absolute current $N\langle j\rangle$ versus the driving strength $\mu$. We compare the nearest neighbor case $\alpha=1000$ (a) (red) with a square root decaying coupling with the distance $\alpha=0.5$ (b) (blue). For the nearest-neighbor case, the system exhibits a NDC as already shown in ref. \cite{driving_clean_chain, negative_differential_conductivity}. Surprisingly, for long-range coupling, the system still acts close to a linear response regime for strongest driving and do not show NDC. The reason for the NDC in chains with strong decaying coupling lies in a spin-blockade which is build up at the boundaries and counteract the inscattering process. Long-range coupling does not show NDC in analogy to the anisotropic Heisenberg-chain with $J_z<J_{x,y}$ \cite{negative_differential_conductivity}.\\
We highlight that for nearest-neighbor coupling there is a crossing of the curves at $\mu^{diff} \approx 0.6$, meaning that the absolute current is the same for all system sizes at that driving strength. $\mu^{diff}$ marks a driving strength, where the system obeys a phenomenological transport law $j = D \nabla \sigma^z$ with $D$ independent of the system size signifying diffusive transport. Note, that the intersection is only at a specific driving $\mu^{diff}$ for $N \ge 7$.  We believe that for smaller systems, finite size effects dominate due to a smaller spin-blockade. For $\mu<\mu^{diff}$, the absolute current increases with the system size $N$ while for  $\mu>\mu^{diff}$ it decreases with $N$. For long range coupling the absolute current is always increasing with $N$, also for maximal $\mu$ and shows linear response to all driving strengths.\\ 
\begin{figure}
\includegraphics[width=0.48\textwidth]{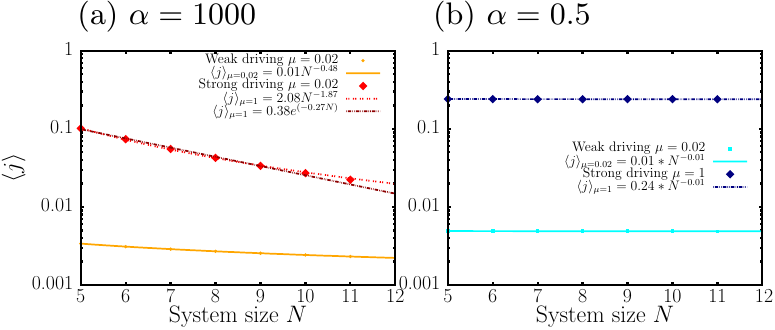}
\caption{Relative current $\langle j\rangle$ versus system size $N$ for weak ($\mu=0.02$) and strong ($\mu=1$) external driving. (a) Nearest-neighbor coupling: For weak driving we obtain $\gamma=0.48$ which is close to the known value $\gamma=0.5$ \cite{PhysRevLett.106.220601} with superdiffusive transport. For strongest driving the transport changes to either subdiffusive transport ($\gamma=1.87$) or an exponential decay. (b) Long-range coupling: In contrast to the nearest neighbor scenario, the transport for long-range coupling is independent of the external drive. For both cases the transport is nearly ballistic with $\gamma=0.01$.}
\label{figure:scaling}
\end{figure}
In order to quantify the transport, we investigate now the relative current $\langle j \rangle$ versus the system size $N$  which we show in Fig. \ref{figure:scaling}. We assume the relative current to scale as $\langle j\rangle \sim 1/N^\gamma$, where the power-law exponent $\gamma$ differs between the transport scenario. $\gamma=1$ we call diffusive transport what is the case at $\mu^{diff}$. Any $\gamma<1$ we call superdiffusive with the case $\gamma=0$ signifying ballistic transport. If $\gamma >1$ the transport is called subdiffusive. We distinguish between the two relevant cases here, weak and strong driving in the following.

\subsection{Weak driving}
Analyzing the scaling for the weak driving regime with $\mu=0.02$ and system sizes up to $N=11$ (Fig. \ref{figure:scaling}), we obtain $\langle j \rangle_{\mu=0.02} \sim 1/N^{0.48}$ for the nearest-neighbor case. This value agrees with $\gamma=0.5$ of Ref. \cite{PhysRevLett.106.220601, diffusive/subdiffusive_Lindblad_chain} obtained for system sizes up to $N=250$. The transport of the clean case is therefore superdiffusive for nearest-neighbor coupling and weak driving within the linear response regime.\\
For the long-range scenario for the clean case, we obtain $\langle j \rangle_{\mu=0.02} \sim 1/N^{0.01}$ which is nearly ballistic transport. Thus, nearest-neighbor coupling and long-range coupling show already a difference in the transport for weak driving. 
\subsection{Maximal driving}

Leaving the linear response regime by choosing strong driving, the nearest-neighbor scenario exhibits NDC. As mentioned before, the intersection at $\mu^{diff} \approx 0.6$ in Fig. \ref{figure:driving} signifies diffusive transport. Increasing $\mu$, the power law exponent changes to $\gamma>1$. For maximal driving we obtain $\langle j \rangle_{\mu=1} \sim 1/N^{1.87}$ which signifies subdiffusive transport. In Fig. \ref{figure:scaling}(a) we also show an exponential fit of the data which would indicate an insulating system. We remark that in Ref. \cite{driving_clean_chain} an exponential fit was more adequate. Our data supports more a power law scaling but for our system sizes finite size effects still are non-negligible. Nevertheless, our purpose is to highlight the difference between nearest-neighbor and long-range coupling which is remarkably: For long-range coupling we obtain $\langle j \rangle_{\mu=1} \sim 1/N^{0.01}$ which is the same as for weak driving with nearly ballistic transport. Therefore, for long-range coupling, the transport is not influenced by the external driving $\mu$. To clarify this robustness, we now investigate the reservoir coupling strength $\Gamma$.\\
\subsection{Reservoir dependency}
By varying the external scattering rates, we show that the transport for long-range coupling remains the same for typical reservoirs characterized with $\Gamma,\mu$ while for nearest-neighbor coupling the transport behavior is changed remarkably.  
 \begin{figure}[h]
\centering
\includegraphics[width=0.48\textwidth]{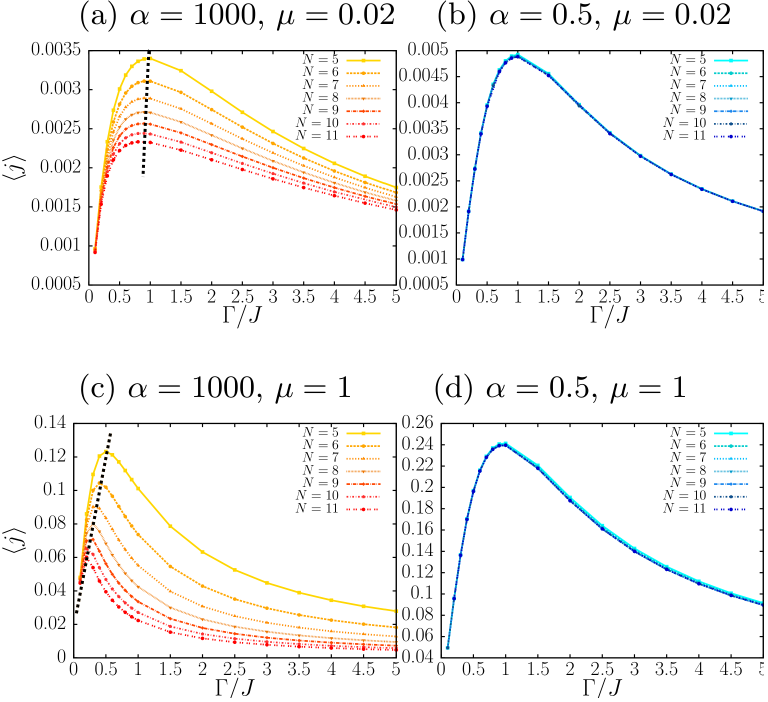}
\caption{Relative current versus the scattering rate $\Gamma$ for weak (a), (b) and strong driving (c), (d). For weak driving, the maximum is at $\Gamma/J=1$ for nearest-neighbor (a) and long-range coupling (b). The current $\langle j\rangle$ decreases with the system size for all $\Gamma$-values for nearest-neighbor coupling while it is the same for all investigated system sizes and $\Gamma$ for long-range coupling. In the far from equilibrium situation the maximum tends to smaller $\Gamma$ for nearest-neighbor coupling and increasing system size (c). For long-range coupling (d) nothing changes but the value of the current.}
\label{figure:gamma}
\end{figure} 
In Fig. \ref{figure:gamma} we again show the two cases: The weak driving regime by comparing nearest-neighbor (a) and long-range coupling (b) as well as the far from equilibrium situation with maximal driving (c) and (d). For weak driving, both, inscattering $\sigma^+$ (spin up) and outscattering $\sigma^-$ (spin down) is present at both sides of the chain with a small bias to spin up (left) or spin down (right) (Eq. \eqref{Lindbladian} and \eqref{Lindbladian_conjugate}). For simplicity we restrict the explanation to the left side, where spin up magnetization ($\Gamma(1+\mu)$) dominates over spin down ($\Gamma(1-\mu)$) in the reservoir. The maximal current is obtained for nearest-neighbor and long-range coupling at reservoir coupling strength $\Gamma/J=1$. The reason is that the spin-flips to the right site dominate over the spin down magnetization of the reservoir $J>\Gamma(1 -\mu)$ while the current is driven by $J<\Gamma(1 +\mu)$. A decreasing of the scattering rate $\Gamma/J<1$ results in lesser inscattering and thus reduces the current to the right side. Increasing of the scattering $\Gamma/J>1$ results also in a higher probability of outscattering (spin down) at the left boundary spin due to $J<\Gamma(1 -\mu)$. As a consequence, only the boundary spins are aligned with respect to the bath magnetization and no further spins are influenced and thus reducing the current. Now let us consider the transport of the nearest-neighbor scenario. At $\Gamma/J=1$ it is superdiffusive with $\gamma\approx 0.5$ as shown before. Either increasing or decreasing $\Gamma$ decreases the power law exponent $\gamma$ as it can be seen qualitatively in Fig. \ref{figure:gamma}(a) that the curves for different system sizes converge. For $\Gamma/J<1$, the influence of the reservoir reduces and thus the polarizations of all spins vanish, becoming independent of the system size. For $\Gamma/J>1$, only the boundary spins polarize with respect to the reservoir due to $J<\Gamma(1 -\mu)$. Therefore, it becomes more and more independent of the system size for increasing $\Gamma$ as the transport of polarization to further spins is reduced. However, for long-range coupling (Fig. \ref{figure:gamma}(b)), all system sizes show the same current, even at the maximum. Independent of the bath magnetization, the transport always remains nearly ballistic. A change of the scattering rate $\Gamma$ only changes the value of the current and not the scaling with the system size.\\
At maximal driving, the left reservoir only contains spin up magnetization  ($2\Gamma\sigma^+$) and no spin down. Any polarization has to be transported to the right side of the chain. Thus, the reason for the decreasing current is different in comparison to the weak driving regime and lies in the spin-blockade effect. A strong alignment of the boundary spin reduces the inscattering probability as the spin is already polarized. For this reason, the maximum is moving to smaller rates $\Gamma/J<1$ at maximal driving (Fig. \ref{figure:gamma}(c)) for nearest-neighbor coupling. Due to the missing outscattering, the spins accumulate up to the central site and a smaller $\Gamma$ is in favor for the current. The number of sites to cross increases with the system size, which is why the maximum moves to smaller $\Gamma/J$ for a higher number of spins. For this reason, the scaling of the current with the system size (cp. Fig. \ref{figure:scaling}(a)) at a specific $\Gamma$ changes dramatically to subdiffusive transport ($\gamma\approx 1.9$) or even exponential decay \cite{driving_clean_chain}. Higher scattering rates $\Gamma/J>1$ intensify the spin-blockade effect, which is why the current decreases, comparable to self-quenching in few emitter lasers. As for the weak driving regime $\Gamma/J\ll 1$ reduces the influence of the reservoir and the current becomes more independent of the system size.\\
In Fig. \ref{figure:scaling}(b) we have shown that for long-range coupling the transport remains nearly ballistic for maximal driving at $\Gamma/J=1$ as it is the case for weak driving. For maximal driving one might expect, that an increasing scattering $\Gamma$ for maximal driving would increase the current as well, as there is maximal potential difference between both reservoirs and no NDC. But on the contrary, even for maximal driving, the maximal current is obtained as well at $\Gamma/J=1$ (Fig. \ref{figure:gamma}(d)). This is surprising, the transport behavior does not change for strong driving due to a missing spin-blockade effect, but still the current for $\Gamma/J>1$ decreases. Also for long-range coupling, the boundary spin polarizes with respect to the bath magnetization. For strongest driving the boundary spin is driven to spin up and counteracts further inscattering. Still, the transport behavior does not change. This effect we explain more detailed in the following section. The maximal driving does not change the system response in comparison to the weak driving (Fig. \ref{figure:gamma}(b)) but for the value of the current. Also for all other investigated $\Gamma$-values, the transport remains nearly ballistic. Therefore, the transport behavior for long-range coupling is completely independent of the potential difference or the reservoir coupling strengths. 
\section{Absence of negative differential conductivity}\label{Absence of NDC}
\begin{figure}[h]
\vspace{0.5cm}
\centering
\includegraphics[width=0.48\textwidth]{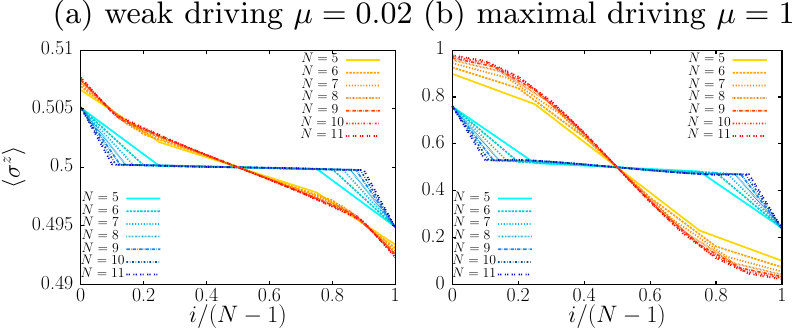}
\caption{Magnetization profile (occupation probability) $\langle \sigma^z_i\rangle$ for the different site indexes normalized by the chain length $i/(N-1)$ for weak (a) and maximal driving (b). Red denotes nearest-neighbor and blue long-range coupling. (a): All values of the single site magnetization are close to $\langle\sigma^z_i\rangle \approx 0.5$, there is a linear decrease of the magnetization from the left side up to the right side of the chain. Only the first and the last site do not show this linear behavior due to the in and outscattering. There is a difference in the linear decrease between long-range and nearest-neighbor coupling defining nearly ballistic and superdiffusive transport. (b) maximal driving: Long-range coupling shows the same magnetization profile  as in the weak driving regime resulting as well in nearly ballistic transport. The magnetization profile for $\alpha =1000$ has changed significantly.  The in-and outscattering influence further sites and the building of a wide spin-blockade takes place which is responsible for the current decrease in Fig. \ref{figure:driving}(a) and is the reason for the change of the transport.}
\label{figure:magnetization}
\end{figure}

Now we investigate in more detail the reason for the absence of NDC in long-range coupled chains and the robustness of the transport behavior against changing reservoir parameters. 
We show the magnetization profile for nearest-neighbor (red) and long range coupling (blue) in Fig. \ref{figure:magnetization}. We again differentiate between the weak driving regime (a) and the far from equilibrium situation with maximal driving (b). Note that the range of the magnetization $\langle \sigma^z_i\rangle$ is different for (a) and (b). 
For weak driving all spins are not aligned in a specific direction. Even the boundary spins are close to $\langle\sigma^z_{0, N-1}\rangle\approx 0.5$ and are only aligned weakly with respect to the bath magnetization. Due to the different bath potentials, a small gradient from left to right is induced. We highlight that already in the weak driving regime this parameter set shows a difference between nearest-neighbor and long-range coupling, leading to superdiffusive transport for $\alpha=1000$ and nearly ballistic transport for $\alpha=0.5$ which we demonstrated in Sec. \ref{characterizing spin-transport}. The reason for the different transport behavior is an increased gradient of the spins close to the center for nearest neighbor coupling. A linear gradient for all spins would indicate diffusive transport as the transport law $j = D \nabla \sigma^z$ is fulfilled with $D$ independent of the position and chain length. We remark that at $\mu^{diff}$ the magnetization profile for nearest-neighbor coupling shows a linear decrease with diffusive transport.\\
In contrast, for strong driving (b), the magnetization profile shows also a qualitative difference between nearest-neighbor and long-range coupling. For $\alpha =0.5$ only the boundary spins are polarized with respect to the bath magnetization while for $\alpha =1000$ also further spins are polarized. This is the spin-blockade which counteracts the inscattering at the left side of the chain, as spin polarizations accumulate up to the centered site. Therefore, the gradient is beyond the linear decrease, which is why the transport changes for strong driving in case of nearest-neighbor coupling. It is clearly visible that for long-range coupling the spins do not polarize except for the boundary spins. Therefore, the spin-blockade is not existent and NDC is absent which is why the transport remains the same for all driving strengths for long-range coupling.
\begin{figure}[h]
\center
\includegraphics[width=0.35\textwidth]{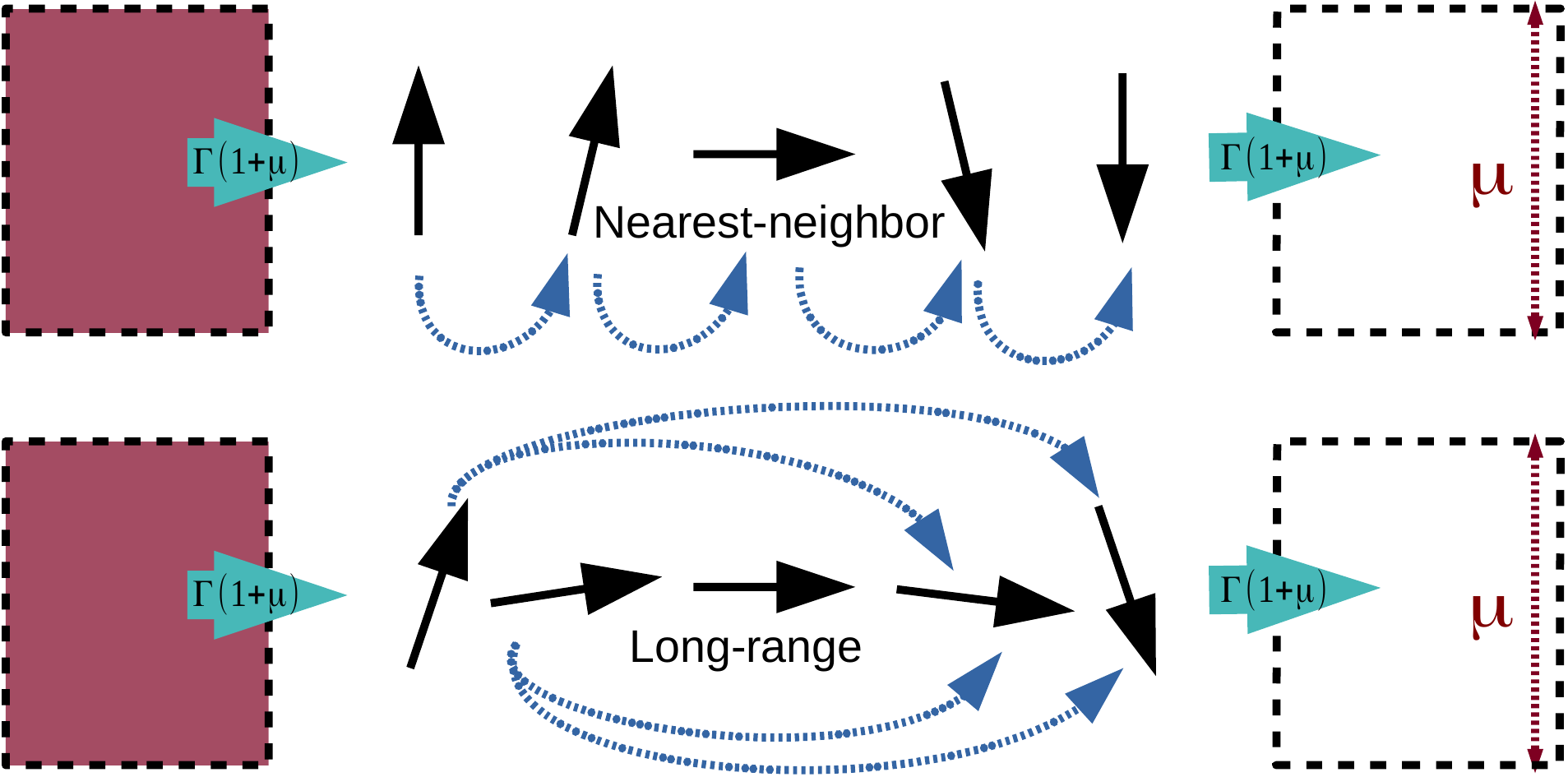}
\caption{Illustration of the spin-blockade for nearest-neighbor and the missing NDC for long-range coupling in case of maximal driving. For nearest-neighbor coupling, the spin polarizations accumulate up to the central site. For long-range coupling, the spins interact with its magnetic counterpart at the other side of the chain and thus reducing the respective polarizations.}
\label{figure:spin-blockade}
\end{figure}
  Long-range coupling enables interactions beyond the central site. The ferromagnetic domain at the left side interacts with its magnetic counterpart at the right side of the chain what we illustrate in Fig. \ref{figure:spin-blockade}. Due to the interaction between the two ferromagnetic domains, the respective polarization on both sides is reduced. However, the boundary spins are still affected by the external reservoirs and polarize, counteracting further inscattering. Therefore, the current also decreases for maximal driving if $\Gamma$ increases (cp. Fig. \ref{figure:gamma}(d)). For nearest-neighbor coupling also further spins are affected by the respective bath polarization and build up a spin-blockade. This bottleneck reduces the current, resulting in NDC. Long-range coupling is robust against accumulation of polarizations and the transport remains nearly ballistic for all driving strengths and all investigated scattering rates $\Gamma$.

\section{Effect of disorder}\label{effect of disorder}
An increase of the external driving results in a change of the transport behavior from diffusive to subdiffusive transport or even exponential decay for nearest neighbor coupling. If disorder is applied, a change of the transport to an insulating exponential decay is of great interest regarding the many-body localization (MBL) transition. In case of weak driving, it was shown by ref. \cite{diffusive/subdiffusive_Lindblad_chain} that the boundary driven nearest-neighbor coupled XXZ-chain contains a transition from diffusive to subdiffusive transport with increasing disorder, before the system exhibits MBL, according to a Griffiths effect \cite{PhysRevLett.114.160401, 1742-5468-2016-6-064010}. We capture this point and show that this transition is highly dependent on the external reservoir for driving beyond the linear response regime. Furthermore, we show that the transport of a long-range coupled chain is not affected by the reservoir and any change in the transport is independent of the external drive.\\
For each data point in the following plots, the disorder is applied randomly at each site with $h_i \in [-h,h] $. For each  disorder realization we evaluate the full density-matrix dynamics until the system converges to a non-equilibrium steady state. We repeat this procedure with new random disorder realizations until the current is well averaged as well as each averaged disorder $\bar h_i \approx 0$. It turns out that for 6000 averages both are well converged. We note that for increasing disorder, the time until the system converges to a non-equilibrium steady state is increasing significantly. Additionally, increasing disorder results in different on-site potentials, wherefore the step size has to be adapted as well.\\ 
Furthermore, for increasing system sizes not only the scaling of the density-matrix with $2^{2N}$ causes numerical effort but rather again an increase of the integration time and an adjustment of the step size. Due to this effort, in this approach we are limited to system sizes up to $N=8$, where we only apply 1000 disorder realizations for each data point. Therefore, our system sizes are significantly smaller than the thermodynamic limit (TDL). 
However, for the purpose of our study, already small system sizes are sufficient. We are interested in the impact of the spin-blockade on the transition from diffusive to subdiffusive transport, even if the exact point of the transition is not the correct value in the TDL. Especially, we compare qualitatively long-range coupling with the nearest-neighbor case, as the spin-blockade is non existent for long-range coupling.\\
In Sec. \ref{characterizing spin-transport}, in the first graph, we have shown the absolute current $N\langle j\rangle$ in Fig. \ref{figure:driving}. For all other plots we showed instead the relative current $\langle j \rangle$, as the transport behavior is obtained from the scaling of the relative current. Now we again focus on the absolute current, because diffusive transport is clearly visible as an intersection of curves with different $N$ as it was the case at $\mu^{diff}$ in Fig. \ref{figure:driving}(a). Similar to $\mu^{diff}$, we find a specific disorder strength $h^{diff}$, where the absolute current does not change for the considered system sizes and an intersection of the curves for different $N$ occurs. Right at $h^{diff}$ the transport obeys a phenomenological transport law $j = D \nabla \sigma^z$ with $D$ independent of the system size. The regime $h<h^{diff}$, where the absolute current increases with the system size is superdiffusive. For $h>h^{diff}$, the absolute current decreases with the system size which is subdiffusive transport.\\ 
In case of a system without boundary reservoirs there exists a second transition at $h^{MBL}\approx 3.7$ \cite{Thermal-eigenstates, Mobility_edge_Heisenberg} where the current scaling with the system size should change from $j\sim 1/N^\gamma$ ($\gamma>1$) to $j\sim exp(-\kappa N)$ which is the many-body localization transition. Here we do not address the question if $h^{MBL}$ exists in such small boundary driven systems but hope these findings might help to characterize $h^{MBL}$ in comparable systems as ours, as long-range coupling shows no bath induced boundary localization (spin-blockade).\\
\begin{figure}[h]
\centering
\includegraphics[width=0.48\textwidth]{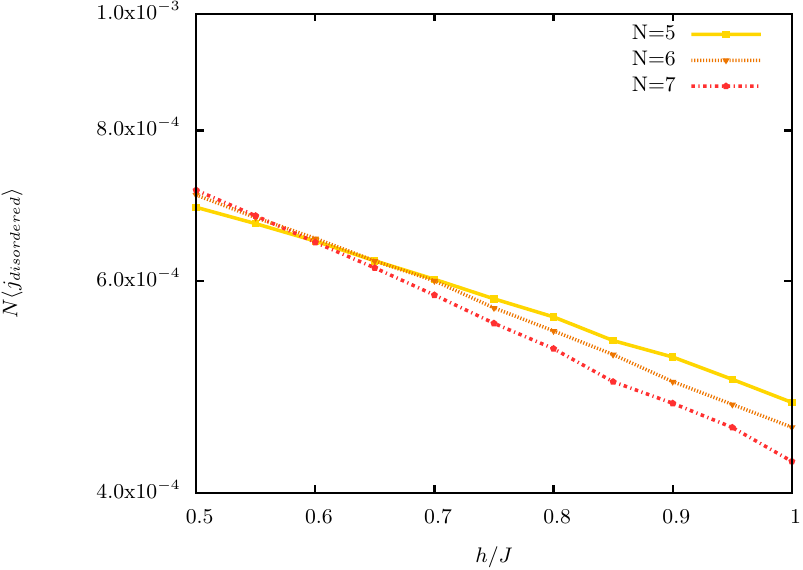}
\caption{Benchmark of the transition to subdiffusive transport visible via an intersection with ref. \cite{diffusive/subdiffusive_Lindblad_chain}. The absolute current $N\langle j\rangle$ does not change for all considered system sizes at a disorder strength $h^{diff}\approx 0.6$. This marks a point with diffusive transport where the system obeys a phenomenological transport law $j =D \nabla \sigma^z$. Before and after $h^{diff}$ the system shows anomalous transport whereby it changes at $h^{diff}$ from superdiffusive to subdiffusve transport. Note that here the disorder at the boundary spins is weaker $h_0, h_{N-1}\in \left[-h/2,h/2\right]$ in order to compare the value of $h^{diff}$ with ref. \cite{diffusive/subdiffusive_Lindblad_chain} .}
\label{figure:intersection_weaker_disorder_at_boundaries}
\end{figure}
We start with a comparison to the findings of ref. \cite{diffusive/subdiffusive_Lindblad_chain}, where they find a transition to subdiffusive transport at $h^{diff}\approx 0.55$. It was shown that there is a critical system size $N^*$ above which finite size effects are absent. We again remark that our investigated system sizes include finite size effect due to $N<N^*$. Still, we can predict with a maximal system size $N=7$, a transition to subdiffusive transport at $h^{diff}\approx 0.6$ what we show in Fig. \ref{figure:intersection_weaker_disorder_at_boundaries}. The resulting disorder strength is close to its value obtained for $N>N^*$ with $h^{diff}\approx 0.55$ \cite{diffusive/subdiffusive_Lindblad_chain}. We note that we are not interested in the exact value of $h^{diff}$ in the TDL, obtained from the current scaling via $ \langle j \rangle \sim 1/N^\gamma$ with $\gamma=1$ at $h^{diff}$. We determine $h^{diff}$ qualitatively from a certain disorder strength where the absolute current $N\langle j\rangle$ does not change for various system sizes $N$. This clarifies that our method to obtain $h^{diff}$ from an intersection of curves can predict a transition relatively close to the value of the TDL for weak driving and nearest-neighbor coupling. In contrast to ref. \cite{diffusive/subdiffusive_Lindblad_chain} we are investigating also the far from equilibrium current. We have shown in Sec. \ref{characterizing spin-transport} that for nearest-neighbor coupling, the chain exhibits NDC resulting in subdiffusive transport already for zero disorder, whereas strong driving does not influence the transport within a long-range coupled chain. Although, we are studying small chains, greater systems show an even stronger NDC for nearest-neighbor coupling (Fig. \ref{figure:driving}), favoring subdiffusive transport for increasing $N$. \\ 
 \begin{figure}[h]
\centering
\includegraphics[width=0.48\textwidth]{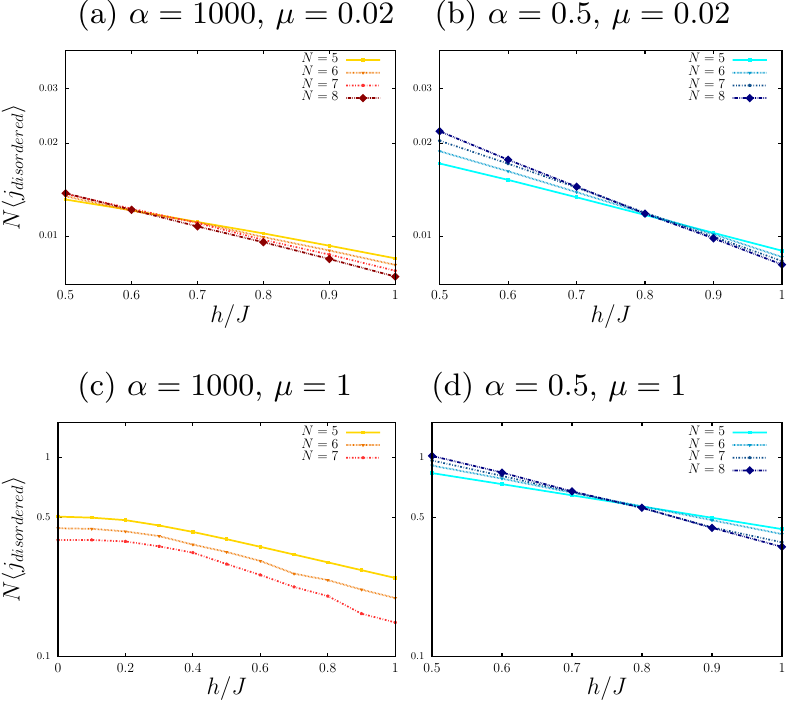}
\caption{Comparison between nearest-neighbor and long-range coupling for weak and strong driving. For weak driving, both nearest-neighbor (a) and long-range coupling (b) show diffusive transport at a certain disorder strength. The transition $h^{diff}$ is at a higher disorder strength for long-range coupling. For maximal driving $\mu=1$, the nearest-neighbor scenario shows transport beyond diffusivity up until $h=0$ (c) while for long-range coupling $h^{diff}$ does not change significantly (d). Note, that due to the numerical effort, we excluded $N=8$ in (c) because it becomes already clear that the transport is subdiffsive for all investigated disorder strengths. }
\label{figure:intersection_plots}
\end{figure} 
In Fig. \ref{figure:intersection_plots} we compare the nearest-neighbor case with the long-range coupling in case of disorder. For weak driving, we observe that $h^{diff}$ is shifted to higher disorder strengths in case of long-range coupling (Fig. \ref{figure:intersection_plots} (a) and (b)). The reason that the transition to subdiffusive transport takes place at a higher disorder strength for long-range coupling, is that long-range coupling shows nearly ballistic transport ($\gamma\approx 0.01$) in contrast to superdiffusive transport ($\gamma=0.5$) of the nearest-neighbor case. Thus, a higher disorder strengths is needed in order to suppress the transport for long-range coupling such that it becomes diffusive. If we increase the external driving strength $\mu$ to maximal driving, we see even a greater difference between both couplings: While for the long-range scenario $h^{diff}$ remains at nearly the same value (Fig. \ref{figure:intersection_plots} (d)), the nearest-neighbor coupling shows transport beyond diffusivity for all investigated disorder strengths, even for $h=0$ (Fig. \ref{figure:intersection_plots} (c)). The reason for this different qualitatively behavior lies in the spin-blockade for the nearest neighbor scenario and the absence of NDC for long-range coupling which we have shown in Sec. \ref{Absence of NDC}. \\
These findings prove that if disorder is applied, the transition to subdiffusive transport is dependent on the external reservoirs for nearest-neighbor coupling. Far from equilibrium driving already suppresses the current due to a spin-blockade effect and it becomes difficult to unravel the effect of disorder from the spin-blockade. The transport for long-range coupling is unaffected by the external reservoir, even for the far from equilibrium situation. Thus, the transition to subdiffusive transport is a disorder effect, independent of the external drive. In our case with a high potential difference of the reservoirs, a possible MBL-transition would be a purely disorder induced effect in case of long-range coupling.

\section{Conclusion}\label{conclusion}
We investigated the isotropic Heisenberg quantum spin-chain with either nearest-neighbor or long-range interaction. By calculating the non-equilibrium steady state current which is induced by two boundary reservoirs at different potentials, we saw that the transport of the long-range coupled chain is independent of the chosen reservoir parameters.  For far from equilibrium driving, the nearest-neighbor scenario shows negative differential conductivity resulting in a change of the transport already for zero disorder with a specific driving $\mu^{diff}\approx0.6$ with diffusive transport. We have shown that for long-range coupled chains, the negative differential conductivity is absent due to the scattering of the two magnetic counterparts beyond the central sites. Long-range coupling still acts close to a linear response regime for the far from equilibrium situation and the transport remains nearly ballistic with a power law exponent $\gamma=0.01$. This is also the case for all investigated system-reservoir interactions.\\
Adding disorder, we observed a specific disorder strength $h^{diff}$ with diffusive transport for both, nearest-neighbor and long range coupling in the weak driving regime.  For the far from equilibrium situation, we observed diffusive transport  for long-range coupling at nearly the same $h^{diff}$. In contrast, the nearest-neighbor case shows subdiffusive transport already for $h=0$ due to a spin-blockade effect. Long-range coupling is robust against the bath induced spin-blockade.\\
Thus, for nearest-neighbor coupling the transport is highly dependent on both, the external driving $\mu$ and the disorder $h$ in a boundary driven Heisenberg model. In order to distinguish many-body localization as an effect of disorder from the spin-blockade, long-range coupling provides a clear understanding of MBL for boundary driven systems as it is robust against far from equilibrium effects.

\section*{Acknowledgments}
We would like to thank Markus Heyl, Andreas Knorr and Nicolas Naumann for fruitful discussions. We gratefully acknowledge the support of the
Deutsche Forschungsgemeinschaft (DFG) through the project B1 of the SFB 910 and by the school of nanophotonics (SFB 787).

\end{document}